\documentclass[twocolumn,showpacs,preprintnumbers]{revtex4}
\usepackage{amssymb}
\usepackage{amsfonts}
\usepackage{amsmath}
\usepackage{graphicx}
\usepackage{dcolumn}
\usepackage{bm}

\begin{document}

\title{Photo-current and photo-voltage oscillations in the
two-dimensional electron system: screening and "anti-screening" of
a potential profile}
\author{S.~I.~Dorozhkin,$^{1,2}$  I.~V.~Pechenezhskiy,$^2$, L.~N.~Pfeiffer,$^3$ K.~W.~West$^3$, V. Umansky,$^4$ K.~von Klitzing$^1$,J.~H.~Smet$^1$}
\affiliation{$^1$Max-Planck-Institut f\"{u}r
Festk\"{o}rperforschung,
 Heisenbergstra\ss e 1, D-70569 Stuttgart, Germany}
 \affiliation{$^2$Institute of Solid State Physics, Chernogolovka,
Moscow district, 142432, Russia}
\affiliation{$^3$Bell Laboratories, Lucent Technologies, Murray
Hill, New Jersey 07974 }
\affiliation{$^4$Department of Physics, Weizmann Institute of Science, 76100 Rehovot, Israel}

\begin{abstract}
We observe in state-of-the-art GaAs based 2D electron systems microwave induced photo-current
and photo-voltage oscillations around zero as a function of the applied magnetic field. The photo-signals
pass zero whenever the microwave frequency is close to a multiple of the cyclotron resonance
frequency. They originate from built-in electric fields due to for
instance band bending at contacts. The oscillations correspond to
a suppression (screening) or an enhancement ("anti-screening") of these
fields by the photo-excited electrons.

\end{abstract}

\pacs{72.20.Fr, 72.20.My, 73.40.Kp}
\date{\today}
\maketitle

The strong current interest in the photo-response of the 2D
electron system (2DES) exposed to microwaves was triggered by the
discovery, at small perpendicular magnetic fields, of microwave
induced oscillations in the
magnetoresistivity~\cite{Zudov1,Ye,Mani,Zudov2} and
magnetoconductivity~\cite{Zudov3} with minima that even drop all
the way down to zero. Numerous theoretical models have been put
forward to account for the oscillatory behavior. The two most
prominent examples are the establishment of an oscillating
non-equilibrium electron energy distribution function
(NEDF)~\cite{Dorozh1,Dmitriev1} and the scattering assisted
displacement (SAD) of photo-excited electrons against or along
the Hall electric field depending on the ratio between the
circular microwave frequency $\omega$ and the electron cyclotron
frequency $\omega_{\rm c}$~\cite{Ryzhii1,Durst}. The second
mechanism was first treated theoretically in 1970 in
Ref.~\cite{Ryzhii1} well before the experimental discovery of the
microwave induced resistance oscillations.
It was predicted that this mechanism also gives rise to an
oscillating photo-current. The main research thrust both
experimentally and theoretically has however been the influence of
the microwave radiation on the sample resistance. A few exceptions
exists where other measurable quantities were addressed such as
for instance the photo-voltage~\cite{Willett,Bykov} and the
electron compressibility~\cite{Vavilov}.

Here, we investigate the photo-signals in two geometries. The
first one contains alloyed contacts at the perimeter (external
contacts) as well as inside (internal contacts) the mesa.  Strong
magneto-oscillations crossing zero are observed in both the
photo-current and voltage when measured between an
internal and an external contact. In the second geometry,
capacitively coupled contacts have been used to pick-up the
oscillating photo-signal and to tune the potential profile with a
gate voltage. The oscillations apparently come about because of
the existence of built-in electric fields within the sample.

Experiments were carried out on samples produced from three
different wafers A, B and C with electron densities $n_{\rm A}=2.9
\times 10^{11} {\rm cm}^{-2}$, $n_{\rm B}=2.7 \times 10^{11} {\rm
cm}^{-2}$, and $n_{\rm C}=2.7 \times 10^{11} {\rm cm}^{-2}$ and
with zero field electron mobilities $\mu_A=17\times 10^6{\rm
cm}^2/{\rm Vs}$, $\mu_B=7\times 10^6{\rm cm}^2/{\rm Vs}$, and
$\mu_C=21\times 10^6{\rm cm}^2/{\rm Vs}$. While samples A and C
were measured in the dark, sample B reached these characteristics
after illumination with a red LED.
Samples A and B (Fig.~1) consisted of a $2.8\ {\rm mm}$ long and
$0.6\ {\rm mm}$ wide Hall bar with three voltage probes on either
side of the bar and source and drain contacts at the ends. In
addition, a set of 9 square shaped ohmic contacts with a side
length of $0.06\ {\rm mm}$ was fabricated within the mesa. These
internal contacts were arranged in a $3 \times 3$ matrix. The
distance in between two rows was equal to $0.15\ {\rm mm}$, while
columns were separated by $0.4\ {\rm mm}$. All contacts were made
by alloying evaporated Ni/Ge/Au. The microwave induced DC
photo-current was measured with a current amplifier and the
photo-voltage with a precision voltmeter. The results were also
verified with a lock-in technique for $100\%$ modulation of the
incident microwave. For these lock-in measurements the
photo-current was converted to a voltage by measuring either
across a small resistor or the secondary coil of a transformer
whose primary winding carried the induced photo-current. The
second geometry in Fig.~1 was fabricated on sample C  and
consisted of a single ohmic contact touching the sample perimeter
and two gates making up a Corbino geometry. The diameter of the
inner circular gate was $1\ {\rm mm}$, while the inner and outer
diameter of the annular gate were $1.5$ and $2\ {\rm mm}$. The
incident radiation with $100\%$ amplitude modulation induces a
displacement (capacitive) current at the modulation frequency due
to charging of the capacitors which  form between the gates and
the 2DES (for a schematic of the measurement configuration see
Fig.~3). This current was measured across both top gates with a
current and lock-in amplifier combination. The samples were placed
in a microwave waveguide with a cross-section of
$6.5\times13\,{\rm mm}^2$ and immersed in liquid $^3{\rm He}$.
Experiments were performed in perpendicular magnetic fields $B$
from 1.5 K down to 0.5 K. For the data presented the estimated
microwave power in the waveguide at the sample location is
$0.1\,{\rm mW}$. We restrict our discussions to data taken on
samples A and C. The results on sample B were qualitatively
similar to those on sample A.

\begin{figure}[tb]
\includegraphics[]{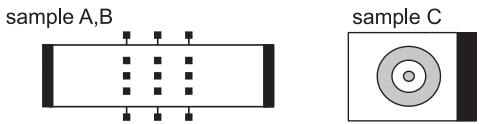}
\caption{Left: Hall bar geometry with internal contacts (sample A,
B).  Right: Corbino geometry with gates. (sample C). Alloyed
contacts are shown in black and gates are grey. }
\end{figure}

\begin{figure}[tb]
\includegraphics[]{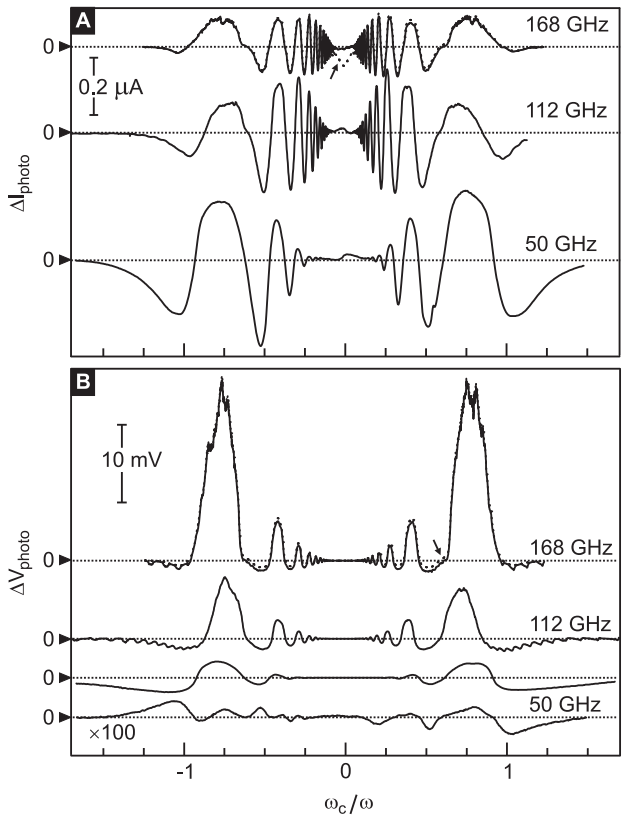}
\caption {The differential DC photo-current $\Delta I_{\rm photo}$
(A) and photo-voltage $\Delta V_{\rm photo}$ (upper three curves
in panel B) measured between an internal and an external contact
for different microwave frequencies $f = \omega/2\pi$ as a
function of $B$. $\omega_{\rm c}/\omega$ was chosen as abscissa.
$\omega_{\rm c}$ is the cyclotron frequency for an electron
effective mass $m^*=0.067m_{\rm e}$. Positive photo-current
values correspond to current flow within the sample from the
external to the internal contact. Curves are offset vertically for
clarity. Horizontal dotted lines mark the zero level. The bare
photo-signals $I_{\rm photo}$ and $V_{\rm photo}$ are marked by a small arrow and
shown as
dashed  lines for $f=168\,{\rm GHz}$. The bottom curve in panel
B has been recorded between two external contacts and was
multiplied by a factor of 100. } \label{Fig2}
\end{figure}

Fig.~2A and B show traces of the DC photo-current and
photo-voltage respectively. They were measured between an internal
and an external contact. To exclude parasitic signals not induced
by the radiation, the plotted data was obtained from a subtraction
of signals recorded in the presence and absence of radiation,
i.e.~ $\Delta I_{\rm photo}=I_{\rm photo}-I_0$ and $\Delta V_{\rm
photo}=V_{\rm photo}-V_0$. The parasitic signals $I_0$ and $V_0$
are small as seen from a comparison of the differential signals
$\Delta I_{\rm photo}$ and $\Delta V_{\rm photo}$ with $I_{\rm
photo}$ and $V_{\rm photo}$. The latter have been included as dashed lines for 168 GHz radiation
(they are marked by a small arrow). The main parasitic effect
appears in the photocurrent near $B=0$. It originates from the
current amplifier when connected to a current source with small
impedance (here the sample at small $B$). The control ac lock-in experiment described previously
yields results (not shown) which are in quantitative agreement
with those presented in Fig.~2. The photo-voltage measured between
any pair of external probes along the
 perimeter is about two orders of magnitude smaller and
quite asymmetric with respect to the $B$-field orientation. An
example for 50 GHz is shown at the bottom of Fig.~2B. Internal
contacts seem superior in detecting the influence of microwaves.
Note the very different shape of the photo-current and
the photo-voltage signals. The photo-current oscillations are
nearly symmetric with respect to zero and exhibit a
non-monotonic dependence of the oscillation amplitude on the
$B$-field at high microwave frequencies. The zeroes in
$\Delta I_{\rm photo}$ and $\Delta V_{\rm photo}$ practically
coincide for signals measured from all 9 internal contacts. The
oscillation amplitude varies by no more than a factor of 1.5 for
the photo-voltage and a factor of 2 for the photo-current when
comparing signals from different internal contacts. The
photo-voltage was also measured between an internal and an
external contact while other internal contacts were short
circuited to the external probe. The photo-voltage amplitude
dropped approximately by a factor of 2 when the nearest neighbor
internal contact was short circuited. This contact is at a
distance of $0.15\ {\rm \mu m}$ from the internal contact used in
the measurement. The photo-voltage is only weakly or not affected
if other internal contacts at a distance of  $0.4\,{\rm mm}$ or
more are short circuited to the external contact. The
photo-current measured when simultaneously connecting two internal
contacts to the current amplifier was close to the sum of currents
produced by the individual contacts. These observations suggest
that the photo-signals arise in the vicinity of each internal
contact.
\begin{figure}[tb]
\includegraphics[]{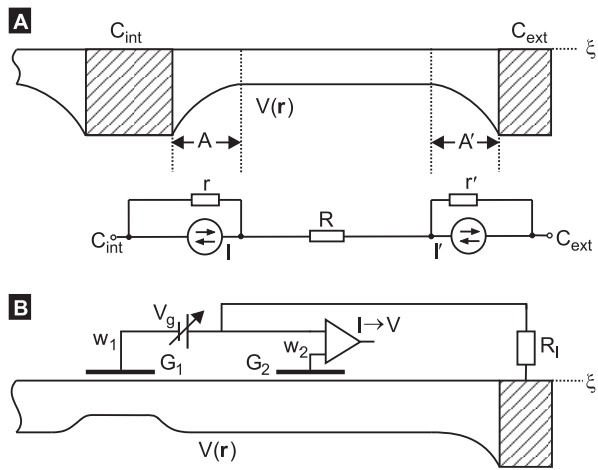}
\caption{(a) Schematic of the potential profile $V({\bold
r})$ in a sample with alloyed internal ($C_{\rm int}$) and
external ($C_{\rm ext}$) contacts. The contact areas are hatched.
Here $\xi$ is the electrochemical potential in  equilibrium. Also shown is the equivalent electrical circuit in the presence of
radiation. (b) Schematic drawing of the potential profile in a
sample with gates $G1$ and $G2$. The measurement circuit to detect
the  capacitive photo-current has been included also.}
\label{Fig3}
\end{figure}

Although both the SAD and NEDF mechanisms provide qualitatively
similar results for the photo-current~\cite{Dorozhkin_tocome} and
are capable of accounting for our experimental findings, below we will
use the easier to visualize SAD picture to explain the experimental data.
This picture has also been instrumental in describing the oscillatory
behavior of the magnetoresistance~\cite{Durst,comment}. When
Landau levels form and a current is imposed through the sample by
applying different electro-chemical potentials at the source and
drain contacts, Landau levels are tilted in a direction
perpendicular to the current flow due to the Hall electric field.
The excitation of an electron to a higher Landau level may then
occur even if the microwave photon energy does not match the
difference between the Landau level energies  $n\,\hbar\omega_{\rm
c}$ ($n = 1,2,\ldots$).  An appropriate scattering assisted
spatial displacement of the electron can simply compensate for the
mismatch in order to fulfill  energy conservation. If
$n-1/2<\omega/\omega_{\rm c}<n$ the electron is preferentially
displaced down hill and the magneto-resistance increases. For
$n<\omega/\omega_{\rm c}<n+1/2$, the electron is promoted against
the electric  force  and the magneto-resistance drops. The
photo-current and photo-voltage oscillations can be understood in
a similar manner even though no external current to tilt the
Landau levels is imposed. The same mechanism is active if built-in
electric fields are present. They take over the role of the Hall
electric field. The inevitable band bending in the vicinity of
contacts as schematically depicted in Fig.~3a is accompanied with
built-in electric fields. For a smooth potential  variation
 Landau levels follow this band bending. A spatial displacement of
 photo-excited electrons allows again to overcome the mismatch between
$\hbar \omega$ and $n \hbar \omega_{\rm c}$
and a photo-current will flow in a direction which  is determined
by the built-in electric field and
the $B$-field value, i.e.~the $\omega_{\rm c}/\omega$ ratio.
The equivalent circuit for this photodiode-like behavior is shown
in Fig.~3a. To emphasize that the photo-current can flow in either
direction despite the given built-in electric field the current
sources have been drawn with a pair of oppositely directed current
arrows. For an open current loop, a difference in the
quasi-electrochemical potential develops across regions A and
A$^\prime$. It drives compensating currents opposite to the
photo-currents to ensure zero net current flow. The shunt
resistance $r$ for the internal contact and $r^{\prime}$ for the
external contact introduce this compensating current in the
equivalent circuit. For photo-current flow against the
built-in electrical field, the generated displacement of electrons
effectively enhances this electrical field and hence this field is
``anti-screened''. This may lead to an instability and a domain
pattern may develop as discussed
in~\cite{Andreev,Halperin1}. For  photo-current  flow along the
electric field the field is suppressed, i.e.~screened, instead. We
assume for simplicity that far away from the contacts there are no
built-in electric fields, so that this area may in the circuit
simply be lumped into a resistor $R$. A pivotal distinction
between internal and external contacts is their shunt resistance
value. In general $r \gg r^\prime$. Since an internal contact is
entirely surrounded by the 2DES, the shunt resistance is Corbino
like and proportional to $\sigma_{\rm xx}^{-1}\approx \rho_{\rm
xy}^2/\rho_{\rm xx}$. Near an external contact however, the shunt
resistance is proportional to $\rho_{\rm xx}$, which under the
experimental conditions is much smaller than $\sigma_{\rm
xx}^{-1}$. This explains why the role of external contacts is only
minor and why a
net photo-current is observed in a closed circuit formed between
external and internal contacts. It follows from the equivalent
circuit that the photo-current $I_{\rm photo}$ and the
photo-voltage $V_{\rm photo}$ measured between contacts $C_{\rm
int}$ and $C_{\rm ext}$ are equal to $(Ir-I^\prime
r^\prime)/(R+r+r^\prime)$ and $Ir-I^\prime r^\prime$,
respectively. Hence, they
are proportional: $V_{\rm photo}=I_{\rm photo}(R+r+r^\prime)$. The
proportionality coefficient is  just the two-point resistance
$R+r+r^\prime$ between both contacts. This resistance can also be
measured independently. It is displayed in Fig.~4 together with
$\Delta V_{\rm photo}/\Delta I_{\rm photo}$. The agreement provides additional support
for the proposed scenario. It also explains the shape difference between the
photo-voltage and photo-current signals.
If it is accepted that band bending at the internal contact occurs
as in Fig.~3a and is dominant, also the sign of the photo-current
is in accordance with the above explanation.
The photo-current will flow within the sample
from the internal  to the  external contact for magnetic fields
above the B-field of zero photo-response ($\omega/\omega_{\rm
c}\lesssim n$) and in the opposite direction for fields below the
location of zero photo-response ($\omega/\omega_{\rm c}\gtrsim
n$). We note that a vanishing photo-response is predicted by all
theories at exactly $\omega/\omega_{\rm c}=n$, but in the reported
magnetoresistance data, the oscillation nodes appear
at slightly  lower  $B$-fields than expected for the reported
GaAs electron effective mass of $0.067\ m_{\rm e}$ (for more
details see Refs.~\cite{Mani2,Zudov4}). This discrepancy with
theory remains unexplained, although it could be eliminated when
assuming a smaller effective mass~\cite{Zudov4}. A similar
observation holds for the photo-effects discussed here.

\begin{figure}[tbh!]
\includegraphics[]{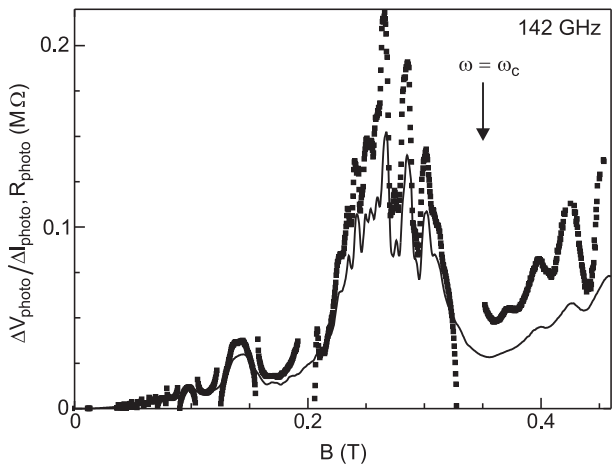}
\caption{Comparison of the 2-point photo-resistance $R_{\rm photo}$ (solid line)
measured with a lock-in technique by driving a small sinusoidal current (137 Hz)
through the sample and $\Delta V_{\rm photo}/\Delta I_{\rm photo}$
calculated from data similar to what is shown in Fig.~2 (symbols). All quantities
were measured between the same pair of contacts and for the same
radiation conditions. The singularities in $\Delta V_{\rm
photo}/\Delta I_{\rm photo}$ are an artefact from the division
through very small values of $\Delta I_{\rm photo}$.} \label{Fig4}
\end{figure}

To further corroborate that built-in electric fields are responsible
for the photo-current and photo-voltage oscillations we have carried out
additional experiments in which the potential profile within
 the sample can be tuned with the help of gates. The geometry is shown in
 the right panel of Fig.~1 and in Fig.~3b.
 A voltage $V_{\rm g}$ is applied
 either to gate $G_1$, while gate $G_{2}$ and the ohmic contact are grounded or vice versa.
 The gates are also used to detect
the capacitive alternating photo-current generated by a train of
microwave pulses at a frequency of 1 kHz and with a duty cycle of
$50\%$. The experiment in essence monitors the redistribution of
charges underneath the gates brought about by the radiation. The
large resistor $R_{\rm l}$ in the measurement circuit of Fig.~3b
suppresses photo-current flow through the alloyed contact, so that
the dominant current in the sample flows between gated
areas.  Experimental results for different values of $V_{\rm g}$
are plotted in Fig.~5 and reveal oscillations similar to those
shown in Fig. 2. At fields larger than 0.2 T, the signals are
strongly affected  by the Shubnikov-de Haas effect and hence we
focus on lower fields. The  oscillations appear already at $V_{\rm
g}=0$ and are either enhanced or suppressed by the application of
a non-zero gate voltage.  In Fig.~5A signals are compared for
$V_{\rm g}=0$ (black curve) and $V_{\rm g}=-800\ {\rm mV}$ (blue
dotted curve). At $V_{\rm g}=-800\ {\rm mV}$ the  oscillations
are  enhanced significantly. Positive values of $V_{\rm g}$ in
the same configuration would cause leakage through the gate.
Instead we may interchange the role of gate $G_{1}$ and $G_{2}$
(i.e.~connect wire $w_1$ to $G_2$ and wire $w_2$ to gate $G_1$ in
Fig.~3b). The oscillations in the capacitive photo-current are now
suppressed when $V_{\rm g}$ drops from 0 (black curve) to $-800\
{\rm mV}$ (blue dotted line). Note that for this modified
configuration the photo-current has the opposite sign, i.e.~maxima
turned into minima and vice versa. The influence of the negative
gate voltage agrees with the phase of the oscillations at $V_{\rm
g}=0$. For $V_{\rm g} = 0$ the data imply that there is a net band bending between
the gated areas which produces
a photocurrent in the sample flowing from the $G_1$-area to the
$G_2$-area when $\omega\gtrsim n\omega_c$ and from the $G_2$-area to
the $G_1$-area when  $\omega\lesssim n \omega_c$.
The application of a negative voltage to gate $G_1$ should then
enhance this band bending  and the amplitude of the microwave
induced oscillations should change as observed in Fig.~5A.
Conversely, a negative gate voltage applied to gate G2
diminishes the net band bending and the oscillation amplitude
should drop. This is indeed born out in the experiment of Fig.~5B.
These results show that the existence of alloyed contacts
is not crucial for the observation of the oscillating photo-signals.
\begin{figure}[tbh!]
\includegraphics[]{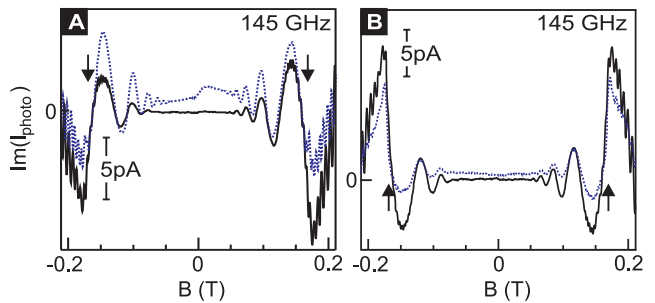}
\caption{The ac current component ${\rm Im}(I_{\rm photo})$
(shifted by 90$^{\o}$ relative to the phase of the microwave
pulses with 1 kHz frequency) versus $B$-field. Black
solid lines and dotted blue curves were measured at voltages
$V_{\rm g}=0$ and $V_{\rm g}=-800\,{\rm mV}$, respectively. The
measurement configuration for the data in panel A is as shown in
Fig.~3(b). For the data in  panel B wire $w_1$ was
connected to gate $G_2$ and wire $w_2$ to gate $G_1$. Positive
current in panel A corresponds to current flow  within the
sample  from the $G_1$-area  to the $G_2$-area, while for panel B the flow
would be from $G_2$ to $G_1$. Arrows mark $\omega=2 \omega_{\rm
c}$} \label{Fig5}
\end{figure}

In summary, incident microwave radiation induces photo-current and
photo-voltage oscillations as a function of the applied $B$-field.
By modifying the local electrostatic potential the oscillation
amplitude can be controlled. The oscillations around zero current
or voltage can be understood at a qualitative level assuming
built-in electric fields and the scattering assisted displacement of photo-excited
electrons~\cite{Ryzhii1}. A quantitative study based on the SAD mechanism or the NEDF mechanism, which can also qualitatively explain the photo-signals, is called for.
The different signs of the photo-signals reflect either the suppression or
enhancement of the built-in electric fields.

The authors acknowledge financial support from INTAS, RFBR (SID
and IVP) the GIF, the BMBF and the DFG.


\begin{references}
\bibitem{comment} Note that the mechanism considered in Ref.~\cite{Ryzhii1} is only one possible
contribution to the photo-response in the magnetoresistance. The
dominant contribution is the non-equilibrium occupation of
electronic states~\cite{Dmitriev1,Dorozhkin2}.
\end{references}
\end{document}